# Conformational Rearrangements in Thin Films of Polydimethylsiloxane Melt

Guennadi Evmenenko,[*] Haiding Mo, Sumit Kewalramani, and Pulak Dutta

*Department of Physics and Astronomy, Northwestern University, Evanston, Illinois 60208-3112*

*To whom correspondence should be addressed:

G. Evmenenko, Department of Physics and Astronomy, Northwestern University, 2145 Sheridan Road, Evanston, IL 60208-3112 USA

Phone: 847-491-3477; Fax: 847-491-9982

E-mail: g-evmenenko@northwestern.edu

**Abstract**

Synchrotron X-ray reflectivity (XRR) confirms the formation of a quasi-immobilized layer in thin films of polydimethylsiloxane (PDMS) melts near silica surfaces. This layer (40-60 Å) has a lower density than the bulk value, and its thickness varies slightly with PDMS molecular weight. Formation of this layer is very rapid for PDMS melts with low molecular weights (below entanglement limit for these molecules) but takes 5-10 hours for higher molecular weights (close to and above their entanglement value).

**Introduction**

Restricted geometries diminish the number of possible molecular configurations in liquids and give rise to ordered structures.[1] For thin films of oligomers and polymers, the region adjacent to the substrate is very sensitive to the shapes and chemical structures of the liquid molecules. Further, depending upon the surface specific interactions and inherent flexibility of the molecular backbones, the molecules may be positionally aligned (layered) at the interface and/or change their shapes to form extended, flat conformations at the interface.[2] Although for short unentangled chains of a polymer liquid (melt) embedded between two parallel impenetrable walls, a surface orientation of the chains parallel to the substrates is predicted[3], recent small-angle neutron scattering experiments show either unperturbed gaussian[4] or distorted conformation[5] of polymeric melt molecules close to an impenetrable interface compared to the bulk. For complex molecules of non-polar van der Waals liquids, the region adjacent to the silicon substrate has a more complicated organization of molecules with a lower density than in bulk.[6] In the case of complex molecules such as elastomers, formation of a gel-like layer of lower density has been confirmed by neutron reflectivity.[7]

In this paper, we describe our structural studies of molecular arrangements in thin films of polydimethylsiloxane (PDMS) melts near solid substrate surfaces. PDMS is a homopolymer with repeat units consisting of $-(O-Si\ (CH_3)_2)-$. A fundamental property of polysiloxane family is the highly pronounced inherent conformational flexibility of the thin main-chain backbones, which results in the high mobility of thin segments and entire molecules. In our previous studies of thin films of low molecular weight PDMS ($M_w <$ 2000 g/mol) at silica interfaces, we had confirmed that molecular layering was induced.[8]

In this case, the interaction between the Si-O groups of PDMS molecules and silica surface is believed to be very weakly adhesive via van der Waals' bonds.[9] Electronic structure studies have previously determined that the dominant adhesive interaction is between the H atoms on PDMS methyl groups and the O atoms in the surface hydroxyl groups.[10] These interactions cause more stable physical bonding between the macromolecules and the hydroxylated silicon surface and lead to flat conformational structures for higher PDMS molecules perpendicular to the substrate. Some evidence of such ordering has been obtained for PDMS molecules with a molecular weight of 2000g/mole.[8] It should be noted that arrangements at the interface are governed by an interplay between the energetics of surface-specific interactions and the entropic factors associated with molecular conformations. In fact, recent NMR studies on higher molecular weight ($M_w = 1.6 \times 10^4$ g/mol) PDMS molecules show uniaxial chain segment ordering in polymer films, even in the absence of surface specific interactions.[11] Also, using only relatively simple physical considerations based on a self-consistent-field lattice model,[12] it was shown for a model polymer melt that (1) there are oscillations of segment potential of about 4 layers for chain molecules in the interfacial region near the wall, and (2) the fraction of molecules with flat conformation increases towards the wall.

The time scales for structural ordering at the interface are governed by the surface diffusion properties in thin liquid films. The diffusion in thin liquid films near an interface is considerably reduced in a two-dimensional process compared to a three-dimensional process.[13] Friction measurements on PDMS melts show a variety of shear-related property changes as a function of film thickness.[14,15] These effects were attributed to different sliding mechanisms of the physiobsorbed and mobile molecular layers. Also, in thin films the presence of an interfacial "solid-like" layer for PDMS films on silica surface has been observed using quasielastic neutron scattering,[16] nuclear magnetic resonance[17] and electrostriction measurements.[18]

Atomic force microscopy measurements show that the time scales for structural changes in PDMS melts at a solid surface are much longer than in the bulk, and that the formation of an immobilized layer at mica and silica surface typically takes 10 hrs for $M_w$ = 18 000 g/mol PDMS molecules.[19-21] However, the authors could not give a value for the layer thickness, because in atomic force microscopy the zero distance is derived from the straight part of the force curve itself and not from an independent quantity.

In the present work, we have sought to examine geometrical confinement effects, and the time evolution of conformational arrangements in thin films of long-chain polydimethylsiloxane melts at hydroxylated surface of silicon substrate.

## Methods and Materials

PDMS samples (Gelest, Inc. Co.) were used as received. The molecular weights $M_w$ of the samples were 3780 (DMS-T15), 13650 (DMS-T23), and 28000 g/mol (DMS-T31). The substrates (3″×1″×0.1″), silicon (111) with native oxide, were purchased from Semiconductor Processing, Inc. They were cleaned in a strong oxidizer, a mixture of 70% sulfuric acid and 30% hydrogen peroxide (70:30 v/v), for 45 minutes at 90°C, rinsed with copious amount of pure water (18 MΩ-cm), and stored under distilled water before use. Prior to preparing the films, the wafers were removed from the water and blown dry under a stream of nitrogen. We made thin films from dilute solutions of PDMS in hexane (0.5-1.0 wt%). The solution was poured over horizontal silicon substrates, and then the hexane was allowed to evaporate in a closed chamber (to slow the evaporation rate). The liquid films we studied were ~ 1-3 μm thick.

XRR studies were performed at the beam lines X23B of the National Synchrotron Light Source and MRCAT of the Advanced Photon Source using a Huber four-circle diffractometer in the specular reflection mode (i.e., incident angle equal to exit angle). X-rays of energy $E$ = 10 keV ($\lambda$ = 1.24 Å) were used for these measurements. The beam size was 0.20-0.30 mm vertically and 1.0 mm horizontally. The samples were placed under helium during the measurements to reduce background scattering and radiation damage. The experiments were performed at room temperature. The off-specular background was measured and subtracted from the specular counts.

## Results and Discussions

The polymer melt structure can be described as "a heap of spaghetti". Depending upon physico-chemical conditions their packing can be intermediate between complete nematic order and random coil packing. The degree of ordering depends on the local stiffness of polymer chains, which is characterized by the Kuhn length or persistence length. PDMS consists of essentially flexible molecules with persistence length of around 5.3 Å. Such conformational flexibility alone is not enough steric reason for significant alignment of PDMS chain segments in bulk. However, the reduced dimensionality and molecular diffusion at an interface, in combination with weak adhesive interactions between the molecules in the melt and the substrate, can lead to the formation of immobilized layers in thin films of polymer melts near hydroxylated silica surfaces. Our recent studies of liquid films (poly(methylhydro-dimethyl)siloxane copolymers) on pure silica substrate and silica substrates with hydrophilic and hydrophobic organic coatings suggest that the region at the interface is very sensitive to the interaction between the liquid molecules and the substrate surface.[22,23] Increasing the number of Si-H groups interacting with the surface leads to a more pronounced low-density region in the vicinity of the substrate. The changes in the electron density profile are also highly dependent on the film thickness. The low-density region increases both in width and in the magnitude of the density dip for the films of ~70 Å and thicker. We have attempted to resolve such a layer for higher molecular weight PDMS ($M_w$ > 2000 g/mol). In such cases, weak substrate-melt interactions can be enhanced by increased polymer chain length, and such a layer can thus have different electron density near the substrate than in the bulk. Observing this experimentally is not an easy task because the expected differences in density are not high, and only a rigorous optimization of all experimental steps (preparation of films, experimental conditions, etc.) has enabled us to get significant results.

Our earlier studies of very thin (40-90 Å) PDMS liquid films revealed a periodicity of about 10 Å for lowest molecular weight oligomers ($M_w$ < 1000 g/mol). The density oscillations are suppressed upon increasing the molecular weight.[22] An increase of molecular length causes a stronger interpenetration of the chains that suppresses inhomogeneities. Structural studies of PDMS with molecular weight of 5200 g/mol do not show any ordering peaks unless shear is applied; a diffraction peak corresponding to the backbone diameter of the PDMS molecule is visible only after applying shear rates of ~$10^4$ s$^{-1}$ for 1.5 min.[15] This 30-Å layer of PDMS relaxes back to the disordered state in ~3 h as a result of diffusion and re-entanglement of the chains, as indicated by the loss of peak intensity.

It has been confirmed by small-angle neutron scattering studies that the conformation of linear PDMS in bulk agrees well with theoretical predictions for Gaussian random-coil polymers.[24] The calculated unperturbed mean-square radii of gyration of PDMS molecules used in our study are approximately 15.4 Å, 31 Å and 45 Å. However, reduced dimensionality at the interface and interactions between the molecules and the substrate may change the conformation of PDMS chains close to substrate surface.

Figures 1-2 show normalized XRR data (specular reflectivity $R$ divided by the Fresnel relectivity $R_F$ for an ideally flat surface of the substrate) from typical scans of different PDMS films as a function of time after preparation. Since the film thickness varies between 1-3 μm, no "Kiessig fringes"[25] corresponding to interference between reflections from the top and bottom of the films have been observed. However, as one can see from the reflectivity, clear minima appear with time. For lowest molecular weight PDMS molecules (DMS-T15), structural rearrangement and attainment of the immobilized layer occurs approximately 2 hrs after film preparation (Fig. 1). For higher molecular weights it takes much longer times: around 4-6 hrs for DMS-T23 (not shown) and 10-15 hrs for DMS-T31 (Fig. 3). Once formed, the immobilized layer in the near-substrate region becames equilibrated,[26] as reflected by the absence of any changes in XRR data. Such molecular weight dependence and response over long time scales is not unexpected and is a characteristic of numerous other polymer phenomena. Different time scales for molecular rearrangement as a function of molecular weight can be explained by different chain mobility, restricted by different degree of attachment to the surface as well as by entanglement with neighboring chains (an entanglement limit of PDMS is ~ 12 000 g/mol).[27]

Additional structural information can be obtained by fitting the XRR curves to a model. In order to find the specific electron density profile, we have fitted the reflectivity curves using the Gaussian-step model. We have assumed that the liquid film consists of a region of uniform bulk density and a immobilized layer close to substrate with variable electron density, thickness and interfacial widths. We have also accounted for silicon and silicon oxide substrate layers. The normalized reflectivity can be written as

$$\frac{R(q_z)}{R_F(q_z)} = \left| \sum_{i=0}^{N} \frac{(\rho_i - \rho_{i+1})}{\rho_0} e^{-iq_z D_i} e^{-q_z^2 \sigma_{i+1}^2/2} \right|^2, \tag{1}$$

where $q_z = (4\pi/\lambda) \sin \theta$; $\lambda$ is the wavelength of the X-ray beam; $\theta$ is the incident angle; $N = 2$; $\rho_0$, $\rho_1$, $\rho_2$ and $\rho_3$ are respectively the electron densities of the substrate (= $\rho_{Si}$), silicon oxide, immobilized layer and PDMS bulk; $D_i = \sum_{j=1}^{i} T_j$ is the distance from the substrate surface to the $i^{th}$ interface; $T_i$ is the thickness of the $i^{th}$ layer and $\sigma_i$ are Gaussian broadened interfaces. The fitting parameters were the thicknesses of the silicon oxide and immobilized layers, the electron density of immobilized layer,[28] and the root-mean-square width of each interface. The fits were performed only for data for which $q > 2q_c$, and the best fits are shown by solid lines in Figures 1-2. The calculated typical electron density profiles for thin films of PDMS melt (samples DMS-T15 and DMS-T31) is shown in Figure 3. A ~40 Å thick reduced density region is formed at the interface between the hydroxylated surface of the silica substrate and PDMS bulk. The electron density of this layer is 5-7% less than the bulk density. A number of fresh samples were prepared and measured by XRR. Analysis of these data allows us to conclude that thickness of immobilized layer depends on molecular weight of PDMS (length of polymer chain) and varies from 40-45 Å (for the shortest chains, DMS-T15 sample), 50-60 Å (DMS-T23) and 60-70 Å (for the longest molecules studied, DMS-T31). If we compare these values with the radius of gyration, we can see that thickness of this layer is approximately $3R_g$ for DMS-T15 and decreases to $1.5R_g$ for DMS-T31. Unfortunately, the fitting process is not precise enough to allow quantitative conclusions about how the electron density of the immobilized layer varies as a function of PDMS molecular weight. We only can claim the presence of a trend: with increasing molecular weight the electron density of this layer approaches the density of bulk PDMS.

Why is the region with reduced density formed in deposited PDMS films and why do the time scales for formation vary with molecular weight? As already mentioned, PDMS molecules have attractive interactions with the hydroxylated surface of the silicon substrate. This should keep the chains at the substrate surface and lead to increased (at least for size of PDMS strand 8-10 Å) or oscillated (with the same scale) density. Presumably due to chain entanglement, we do not see any manifestations of layering or other ordering at the film-substrate interface (see, for example, refs. 8 and 15 for a detailed discussion of ordering in thin films of short-chain PDMS as a function of molecular weight or applied shear). We speculate that such a picture is valid only for very short times after film deposition. This state is kinetically metastable. Since in pure bulk polymers the average shape of chains in contact with the surface is determined only by entropic factors, the reduced configurational entropy of confined polymer chains can only increase by (partially) leaving the surface and assuming a random coil configuration.[29]

The loss of configurational freedom (relative to the bulk) by polymers in proximity to the surface produces an "entropic repulsion" between the polymer and the wall[30] and as a consequence, the appearance of a depletion layer at the interface.[31,32] Such rearrangement can take a long time and result in compression of these random coils and a reduced density of polymer melts at interfaces. Polymer density profiles are dependent on chain lengths and polymer architectures, and are more inhomogeneous for longer- or branched-chain molecules.[32] At the same time, recent Monte Carlo simulations of confined polymer melts (without considering energetical contacts) show controversially that regardless of melt density and chain length, the segment densities can be described by a universal curve with a considerably lower density for the layers closest to the confining surfaces.[33] But we can not eliminate interaction of polymers with a surface; this is one of the most important factors that influence the static and dynamic properties of confined polymers. De Gennes has mentioned the importance of "pinning" of polymer molecules to the solid surfaces for the existence of long-range repulsive forces and the formation of an immobilized layer either by specific interactions with the surface or by solidification of the liquid very close to the surface.[30] Even polymers of low molecular weight can be pinned for a significant time to a solid surface due only to topological reasons, as was shown in ref. 34. The introduction of van der Waals polymer-polymer interactions can also considerably affect the behavior of the polymer melt near the wall (and reduce the density dip even further) as was shown in a Landau-Ginsburg free energy functional theory developed for describing a polymer melt at a hard wall.[32] So there are different energetic and entropic factors which have the effect of immobilizing the polymer liquid, and the interplay of these factors can be a reason for the observed differences in thickness of the immobilized layers and the incubation time for different molecular weight PDMS melts.

**Conclusions**: Evidence of conformational rearrangement of molecules at solid surface is obtained by synchrotron X-ray reflectivity for thin films of PDMS melts deposited on hydroxylated silicon substrates. Immobilized layers of reduced electron density are formed by confined polymer chains. The thickness of the layer varies slightly with the chain length. The major difference however is in the time scales of formation of these layers: attainment of the equilibrated immobilized state occurs very fast for shorter-chain molecules and much more slowly (between 5-10 hours) for higher molecular weight PDMS (greater than the entanglement limit of PDMS). Such differences in behavior can be attributed to the interplay of energetic factors (imposed by specific interactions of melt molecules with the substrate surface) and entropic factors (excluded volume effect and reduced dimensionality at the melt-solid interface).

**Acknowledgements**. This work was supported by the US National Science Foundation under grant no. DMR-0305494. XRR measurements were performed at beam lines X23B of the National Synchrotron Light Source and the MRCAT of the Advanced Photon Source, which are supported by the U.S. Department of Energy.

**FIGURE CAPTIONS**

**Figure 1.** X-ray reflectivity data for thin films of DMS-T15 (3780 g/mol) at different times after film deposition: 15 min, 1, 2 and 15 hours (from the top to the bottom). Solid lines are best fits using Eq. 1. The curves are shifted vertically for clarity.

**Figure 2.** XRR data for thin films of DMS-T31 (28000 g/mol) at different times after film deposition: 1, 5, 10, 15 and 25 hours (from the top to the bottom). Solid lines are best fits. The curves are shifted vertically for clarity.

**Figure 3.** Examples of calculated electron density profiles from fits of XRR data for DMS-T15 (curve 1) and DMS-T31 (curve 2) films after formation of the equilibrated immobilized layers (z is distance from the hydroxylated surface of silicon substrate).

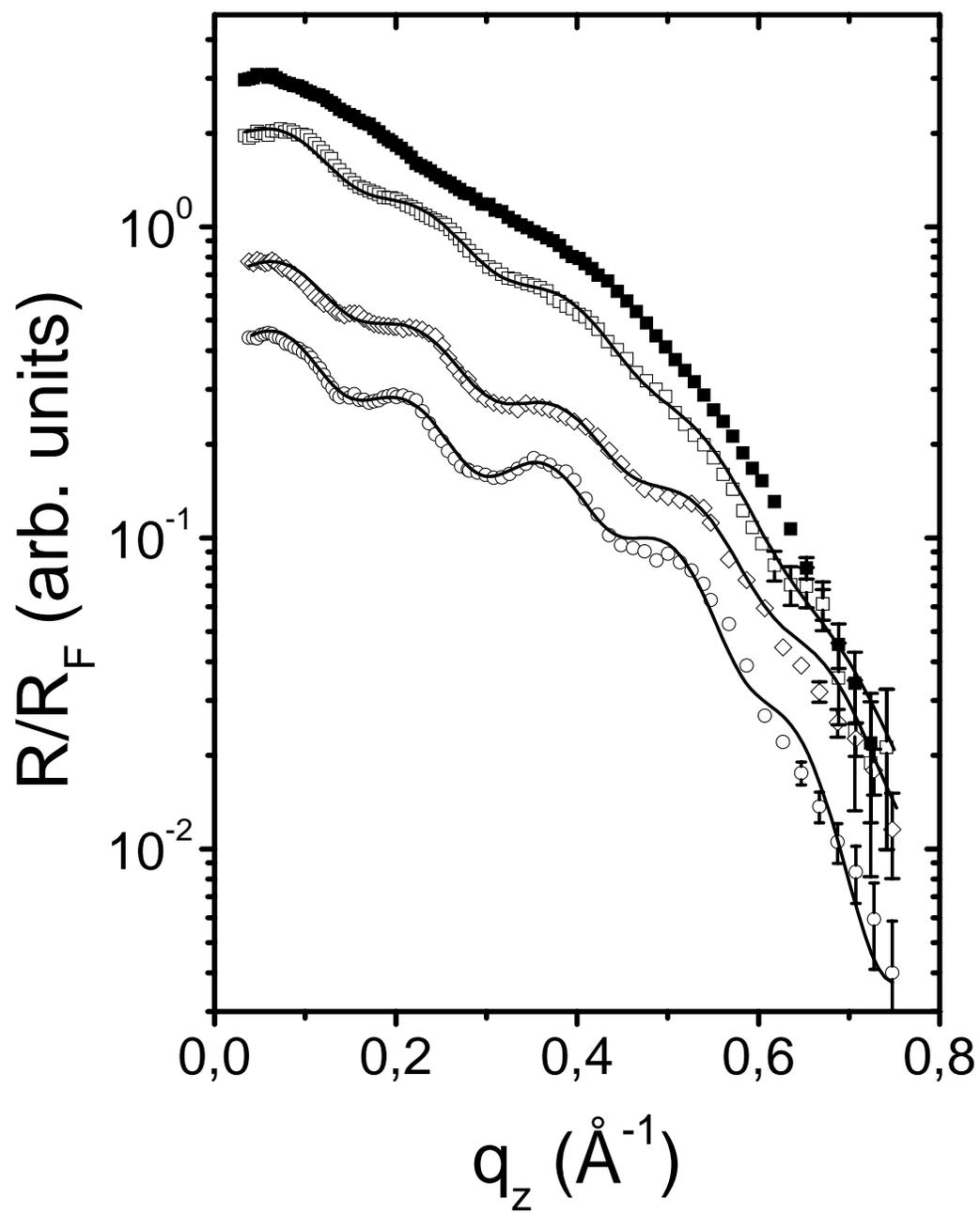

**Fig. 1.**
*Evmenenko et al.*

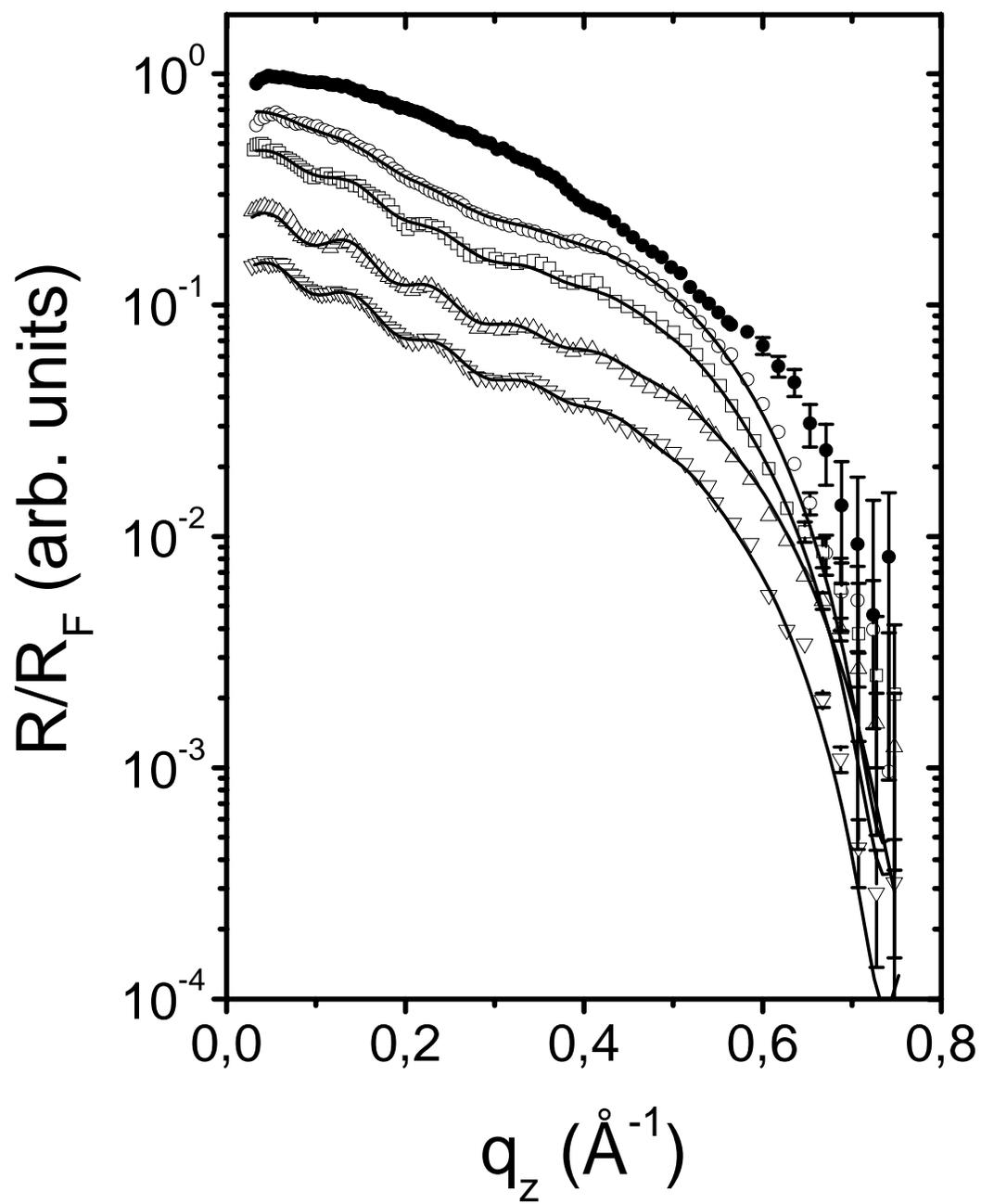

**Fig. 2**
*Evmenenko et al.*

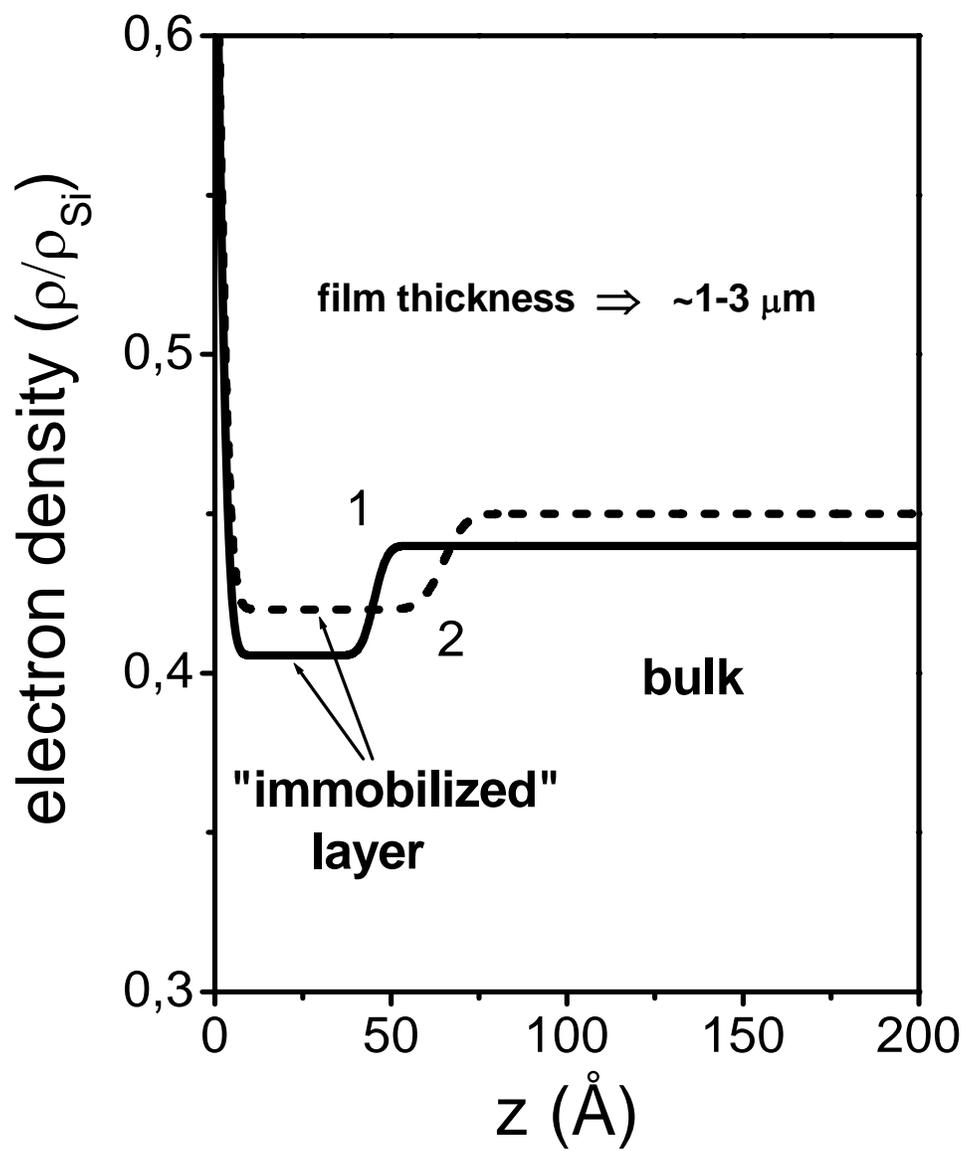

film thickness $\Rightarrow$ ~1-3 μm

**Fig. 3**
*Evmenenko et al.*